\documentstyle[twocolumn,prl,aps,epsf]{revtex}
\tighten
\begin{document}
\draft
\title{Re-entrant Melting in Polydisperse Hard Spheres}
\author{Paul Bartlett}
\address{Department of Chemistry, University of Bath, Bath, BA2 7AY, United 
Kingdom}
\author{Patrick B. Warren}
\address{Unilever Research, Port Sunlight Laboratory, Quarry Road East, 
Bebington, Wirral, L63 3JW, United Kingdom}
\date{8$^{th}$ December, 1998}
\maketitle

\begin{abstract}
The effect of polydispersity on the freezing transition of hard spheres is 
examined within a moment description. At low polydispersities a single 
fluid-to-crystal transition is recovered. With increasing polydispersity we find 
a density above which the crystal melts back into an amorphous phase. The range 
of densities over which the crystalline phase is stable shrinks with increasing 
polydispersity until, at a certain level of polydispersity, the crystal 
disappears completely from the equilibrium phase diagram. The two transitions 
converge to a single point which we identify as the polydisperse analogue of a 
point of equal concentration. At this point, the freezing transition is 
continuous in a thermodynamic sense.
\end{abstract}


Freezing and melting are probably the most common and striking physical changes 
observed in everyday life. All experiments, to date, demonstrate that the 
crystallisation of a simple liquid is a first-order transition, in three 
dimensions. So for instance, the sharp Bragg peaks of the crystal, which reflect 
the long-range spatial modulation of the density $\rho(\bbox{r})$ and which 
distinguish a crystal from a liquid, disappear {\em abruptly\/} as  a crystal 
melts\cite{Chaikin-1263}. This sharp microstructural change is also mirrored by 
discontinuities in the first derivative of the free energy so that 
experimentally, melting is accompanied by a finite density and entropy change.

Although the experimental situation is clear, in an early analysis 
Landau\cite{Landau-1404} argued that, under certain conditions, a crystal can  
transform  {\em continuously\/} into a liquid. In a simple 
Landau-Alexander-McTague theory\cite{Alexander-1310} the excess free energy of 
the crystal (relative to the isotropic liquid) has the following form:
\begin{eqnarray}
\label{landau}
f_{sl}& = & r(T,P)\sum_{\bbox{G}} |n_{\bbox{G}}|^{2} \\
& & - u_{3}(T,P)\sum_{\bbox{G_{1}},\bbox{G_{2}},\bbox{G_{3}}} n_{\bbox{G_{1}}} 
n_{\bbox{G_{2}}} n_{\bbox{G_{3}}} 
\delta_{\bbox{G_{1}}+\bbox{G_{2}}+\bbox{G_{3},0}} 
+ \ldots \nonumber
\end{eqnarray}
where the order parameters $n_{\bbox{G}}$ are the Fourier components of the 
crystal density, $\rho_{s}(\bbox{r}) = \rho_{s} + \delta \rho(\bbox{r})$, at the 
reciprocal lattice vector $\bbox{G}$ ($\rho_{s}$ is the uniform crystal density) 
and the coefficients of the expansion are analytic functions of the temperature 
$T$ and pressure $P$. Eq.~\ref{landau} contains cubic terms because the order 
parameter sets $\{n_{\bbox{G}}\}$ and $\{-n_{\bbox{G}}\}$ describe physically 
distinct crystals with different energies. As a consequence the  freezing 
transition is generally first-order. However since {\em both} $T$ and $P$ can be 
independently varied the possibility exists that $r$ and $u_{3}$ can be made to 
vanish at a  single point in the $T-P$ plane. At the resulting Landau point the 
liquid-solid transition is  continuous in a mean-field description\cite{Note}. 
Landau theory makes two further distinctive predictions. First the Landau point 
must lie at the intersection of, at least, three first-order lines of 
transitions\cite{Landau-1404} which separate the liquid from two conjugate 
crystalline phases, C$_{+}$ and C$_{-}$, with identical symmetry but which 
differ in the sign of $\delta \rho(\bbox{r})$. Second, in three dimensions, 
symmetry considerations should uniquely favour a bcc 
structure\cite{Alexander-1310}. 

In spite of these interesting predictions it is not clear if, in a liquid-solid 
system, the point at which the cubic coefficient $u_{3}$ vanishes is 
experimentally accessible. On the face of it, one of the most promising 
candidates is a system of polydisperse hard spheres where the constituent 
particles have different sizes. The freezing of polydisperse hard spheres has 
been studied extensively in recent 
years\cite{Dickinson-848,Barrat-810,Pusey-829,%
McRae-414,Moriguchi-1232,Vermohlen-858,Bolhuis-824,%
Bartlett-983,Sadr-1165,Phan-1265} motivated, in part, because it is a realistic 
model of a colloidal suspension\cite{Pusey-853}. These studies have focused 
mainly on the effect of size polydispersity $\sigma$, defined as the ratio of 
the standard deviation to the mean of the diameter distribution, upon the 
fluid-solid transition. Calculations have been made using a variety of 
theoretical and computational techniques, for various size distributions, and in 
both two and three dimensions. Yet the picture that has emerged is remarkably 
similar. On increasing $\sigma$, from zero the density discontinuity at the 
transition $\Delta \rho = \rho_{s} - \rho_{l}$ decreases, vanishing altogether 
at a ``terminal'' polydispersity\cite{foot}, $\sigma = \sigma_{t}$, above which 
no liquid-solid transition is found. A number of key questions have however been 
left unanswered. First, why do the densities of the coexisting phases converge 
as $\sigma \rightarrow \sigma_{t}$? If the liquid-solid transition is continuous 
then the singularity at $\sigma_{t}$ must correspond to a Landau point. The 
phase diagram should therefore contain {\em two\/} crystal phases, in 
contradiction with the theoretical work to date. Furthermore while the C$_{+}$ 
crystal has the normal bcc structure with spheres at the cube corners and 
centre, the C$_{-}$ crystal has particles at interstitial sites. The 
unfavourably low packing of the C$_{-}$ crystal ($\phi_{m} \sim 0.20$) makes it 
unlikely that this phase could be important in a dense system. If the vanishing 
density discontinuity at $\sigma_{t}$ is not critical in origin, then what is 
its true nature? And finally, why is the polydisperse phase behavior apparently 
universal? In this letter we reexamine the freezing of polydisperse hard spheres 
using simple mean-field models for the polydisperse crystal and liquid phases. 
{\em Our results suggest that the polydisperse solid-liquid transition at 
$\sigma_{t}$ is not critical.} We show that the vanishing of the density 
discontinuity at the terminal polydispersity is a consequence of a {\em 
re-entrant\/} solid-liquid transition in a polydisperse system. 

Our model consists of $N$ hard sphere particles in a volume $V$, at an overall 
density of $\rho = N/V$. Each particle has a diameter $R$ drawn from a 
distribution $\rho(R)$ so that $\rho = \int dR \rho(R) $. The distribution 
$\rho(R)$ is conveniently characterised by the set of generalised moments $m_{i} 
= \int dR \rho(R) w_{i}(R)$ where the weight function $w_{i}(R) = (R/\bar{R} -
1)^{i}$. The zeroth moment is simply the total number density $\rho$. The 
``shape'' of the diameter distribution, $\tilde{\rho}(R) = \rho(R) / \rho$, is 
taken here for simplicity, as the Schultz distribution, $\tilde{\rho}(R) = 
\gamma^{\alpha} R^{\alpha-1} \exp(-\gamma R) / \Gamma(\alpha)$ with $\alpha = 1 
/\sigma^{2}$ and $\gamma = \alpha /\bar{R}$. The (excess) chemical potential 
$\mu^{ex}(R)$ in a polydisperse system is in general a complex and unknown 
function of the particle size. But with the assumption that there is no critical 
point at $\sigma_{t}$ the excess chemical potential must, first of all, be an 
analytic function of $R$. Formally, $\mu^{ex}(R)$ may be calculated from the 
probability, $W(R)$, for insertion\cite{Widom-849} of a test sphere of diameter 
$R$. At large $R$, the leading term in $\mu^{ex}(R)$ is the $PV$ work required 
to generate a cavity sufficiently large to accommodate the test sphere.  This 
contribution varies as $R^{3}$. Motivated by this we assume that in a  
hard-sphere crystal or fluid $\mu^{ex}(R)$ has the simple analytic form
\begin{eqnarray}
\mu^{ex}(R) &=& -k_{B}T \ln W(R) \nonumber \\
&\approx& \lambda_{0} + \lambda_{1} R + \lambda_{2} R^{2} +\lambda_{3} R^{3}. 
\label{mu}
\end{eqnarray}
where consistency demands that the coefficients $\lambda_{i}$ depend only on the 
four moments $m_{0}, \ldots m_{3}$ of the polydisperse 
distribution\cite{Bartlett-983}. Two of the four unknown coefficients may be 
determined from the known small and large $R$ limits of $W(R)$. This fixes 
$\beta \lambda_{0} = -\ln (1- \phi)$ and $\lambda_{3} = \frac{\pi}{6} P$ with 
$\phi$ the volume fraction and $\beta = 1/k_{B}T$.

Having specified the general form expected for $\mu^{ex}(R)$, we now outline the 
calculation of the size-dependent chemical potential in the crystal. From 
Eq.~\ref{mu} the probability to insert an arbitrary-sized test particle into any 
two hard-sphere systems will be equal if the two distributions have the same 
first four moments\cite{Bartlett-983}. In this sense the two systems may be 
termed ``equivalent''. Since a binary mixture can always be chosen so as to 
match any four moments we look at the ``equivalent'' binary 
substitutionally-disordered crystal, for which simulation data is 
available\cite{Kranendonk-77}. By looking at test particles with sizes equal to 
the two species in the binary mixture, for which the chemical potentials are 
known, the remaining two unknown coefficients ($\lambda_{1}$ and $\lambda_{2}$) 
in the general expression for $\mu^{ex}(R)$ are determined. The resulting  
predictions for the polydisperse crystal has been compared with simulation data 
previously\cite{Bartlett-983}. Agreement is good.

For the polydisperse fluid accurate expression for $\mu^{ex}(R)$ are available. 
We use the approximate BMCSL\cite{Mansoori-852} equation of state which for a 
Schultz distribution has the closed form 
\begin{equation}
\frac{\pi}{6}\beta P_{f}\bar{R}^{3} = \frac{\xi}{1+\sigma^2} + 
\frac{3\xi^{2}}{1+\sigma^2} + (3-\phi)\xi^{3} \label{fluid}
\end{equation}
where $\xi = (\frac{1}{1+\sigma^2})\frac{\phi}{1-\phi}$. The excess free energy 
per particle is found by integrating Eq.\ \ref{fluid}. Differentiation then 
yields an expression for the particle potential $\mu^{ex}(R)$ which is of the 
form of Eq.\ \ref{mu}.

The total polydisperse free energy $f$ (with $f = F/V$) consists of ideal and 
excess terms, $f = f^{id} + f^{ex}$, which depend in a very different manner on 
the distribution $\rho(R)$. The excess free energy, $ f^{ex} = \int dR \rho(R) 
\mu^{ex}(R)$, is a function only of the four moments variables $m_{0}, \ldots 
m_{3}$. The ideal term $\beta f^{id} = \int dR \rho(R) \ln (\rho(R))$, by 
contrast, depends upon the detailed shape of the function $\rho(R)$ so formally, 
at least, the total free energy $f$ resides in an infinite dimensional space. 
Sollich, Cates and Warren\cite{Sollich-1032} have shown that the full 
polydisperse phase diagram can be approximated by replacing the ideal free 
energy by a projected term $\widehat{f^{id}}(\{m_{i}\})$ which includes only 
those contributions that depend on a {\em finite\/} set of moment variables. The 
remaining contributions to the ideal free energy, from those degrees of freedom 
of $\rho(R)$ which can be varied without affecting the selected moments, are 
chosen to minimise the free energy. The power of this approach is that by 
including more moment variables the calculated phase diagram approaches, with 
increasing precision, the actual phase diagram. The position of equilibrium is 
fixed by the equality of the `moment' chemical potentials, $\mu_{i} = \partial 
\widehat{f}/\partial m_{i}$ and the pressure $P$ among all phases with 
$\widehat{f}$ the projected free energy. For polydisperse hard spheres the 
excess moment chemical potentials are simply combinations of the (known) 
coefficients \{$\lambda_{i}$\} in $\mu^{ex}(R)$ (Eq.~\ref{mu}) since 
$\mu(R)=\delta \widehat{f}/\delta \rho(R)=\sum_{i} (\partial 
\widehat{f}/\partial m_{i}) w_{i}(R) = \sum_{i} \mu_{i} w_{i}(R)$. The first two 
ideal moment potentials are\cite{Sollich-1032}, ignoring unimportant factors, 
$\mu^{id}_{0} = \ln \rho - \alpha \ln \bar{R}$ and $\mu^{id}_{1} = -\alpha 
\bar{R}$.

\begin{figure}
\begin{center}
\epsfxsize=2.4in 
\leavevmode 
\epsfbox{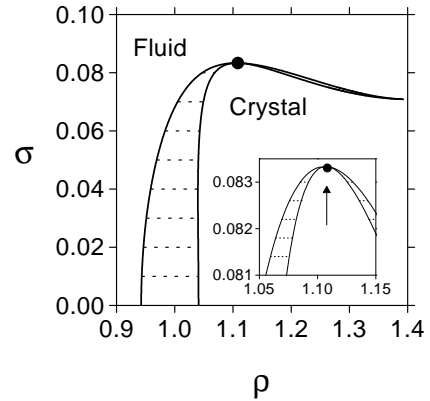}
\end{center}
\caption{Phase diagram of a polydisperse system of hard spheres showing the 
re-entrant freezing transition. The density discontinuity $\Delta \rho = 
\rho_{s} - \rho_{l}$ vanishes at the {\em point of equal concentration\/}, 
marked by the filled circle. The inset figure shows the phase boundaries near 
this point in greater detail.}
\label{fig1}
\end{figure}

In order to understand the qualitative features of the polydisperse transition, 
we consider first the simplest description in which only the lowest moment 
($m_{0}$) is retained in the projected free energy.  In this limit, there is no 
size fractionation so the normalised diameter distribution, $\tilde{\rho}(R)$, 
is fixed and equal in all phases. The location of the fluid-solid transition is 
determined by equating $P$ and $\mu_{0}$, the chemical potential of the 
mean-sized particle, in each of the crystal and fluid phases. In this way we 
obtain the phase diagram of Fig.~\ref{fig1}. At low densities we find, in 
qualitative agreement with previous 
work\cite{Dickinson-848,Barrat-810,Pusey-829,McRae-414,%
Moriguchi-1232,Vermohlen-858,Bartlett-983,Sadr-1165,Phan-1265}, that the density 
discontinuity at freezing $\Delta \rho$ reduces with increasing polydispersity 
and eventually vanishes at the point $\sigma_{t} = 0.0833$ and $\rho_{t} = 
1.111$. However, at high polydispersity, the calculated diagram contains a novel 
feature. For $0.07 \leq \sigma \leq 0.083$ we find a further transition from the 
crystal back to a disordered phase\cite{note1}. The location of this 
polydispersity-induced-melting transition varies sharply with polydispersity. 
The range of densities for which a crystal is found shrinks with increasing 
polydispersity until at $\sigma_{t}$ the crystal of density $\rho_{t}$ 
disappears completely from the phase diagram. At the point 
($\rho_{t}$,$\sigma_{t}$) the line of fluid-to-crystal transitions intersects an 
upper line of crystal-to-amorphous transitions. At all points in the 
($\rho$,$\sigma$) plane the freezing transition remains first-order so the 
singularity at ($\rho_{t}$,$\sigma_{t}$) is equivalent to the {\em point of 
equal concentration\/}\cite{Landau-1238} seen in molecular mixtures and is not a 
critical point -- so providing an answer to the second of our questions.

\begin{figure}
\begin{center}
\epsfxsize=2.4in 
\leavevmode
\epsfbox{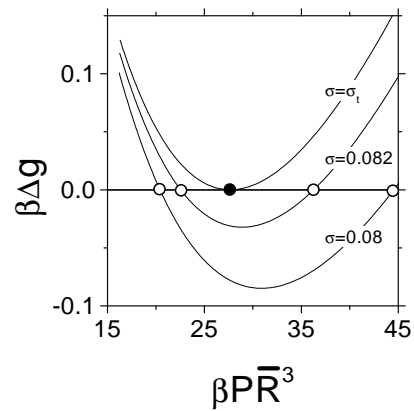}
\end{center}
\caption{The Gibbs free energy difference (per particle) $\Delta g$  between 
crystal and fluid phases as a function of the pressure, for different 
polydispersities. The circles are the first-order fluid/solid transitions. The 
filled circle marks the point of equal concentration.}
\label{fig2}
\end{figure}

We now turn to the vanishing density discontinuity in the vicinity of the point 
of equal concentration. The Gibbs free energy difference $\Delta g = g_{s} - 
g_{l}$ (with $g = G/N$) between the solid and liquid phases as a function of 
pressure for three fixed values of $\sigma$ is shown in Fig.~\ref{fig2}. The 
re-entrant nature of the freezing transition is very evident with a stable 
crystal appearing only in an intermediate range of pressures bounded by the two 
transitions where $\Delta g = 0$. The density change $\Delta \rho$ at the 
liquid-solid transition is given by the slope of the free energy curve at the 
point $\Delta g = 0$ since $\partial \Delta g / \partial \rho = (1/\rho_{s}) - 
(1/\rho_{l})$. Increasing the polydispersity raises the free energy of the solid 
relative to the fluid, displacing the $\Delta g$ curve vertically and as is 
evident from Fig.~\ref{fig2} reduces the density jump at the transition. At the 
terminal polydispersity the solid just touches the fluid curve so the tangent is 
horizontal and $\Delta \rho = 0$. In a system of hard spheres (where the 
internal energy is constant) the condition $\Delta \rho = 0$ necessarily 
requires the entropy change at this point to also vanish. Clearly while the 
underlying microscopic transition remains first-order the first derivatives of 
the thermodynamic potential are continuous at $\sigma_{t}$. A conventional 
classification of this transition, following the ideas of 
Ehrenfest\cite{Stanley}, is clearly inappropriate.

Retaining more moments in the projected free energy allows the possibility of 
different-sized particles to be partitioned between phases. To establish the 
effect of fractionation we have recalculated the phase equilibria with {\em 
two\/} moment variables. The phase diagram, now given by equating $P$ and the 
moment potentials $\mu_{0}$ and $\mu_{1}$ in all phases, is unchanged in 
topology from Fig.~\ref{fig1}. The point of equal concentration is retained 
although shifted slightly to $(\rho_{t},\sigma_{t}) = (1.115,0.0831)$. Hence our 
prediction of a re-entrant freezing transition seems to be robust. The extent of 
fractionation is generally small, although increasing as $\sigma \rightarrow 
\sigma_{t}$, with the larger particles preferentially found in the crystal 
phase.  Details of our calculations are given elsewhere\cite{me}.

The appearance of an equilibrium amorphous phase may be understood simply from 
maximum packing arguments. For uniform-sized spheres the maximum density of a 
randomly packed Bernal glass ($\rho_{rcp} \simeq 1.22$)  is significantly 
smaller than the geometric limit of a close-packed hexagonal of fcc crystal 
($\rho_{cp} = \sqrt{2}$). The greater packing efficiency of the crystal ensures, 
that at high densities, particles have more freedom and so a higher entropy than 
those in the fluid phase\cite{Baus-268}. The stable high density phase of 
uniform hard spheres is therefore crystalline. Polydispersity affects 
crystalline and disordered phases in different ways. In an amorphous phase, 
small particles pack in the cavities between large particles and $\rho_{rcp}$ 
increases with $\sigma$ while the constrained environment of a fixed repeating 
unit cell causes the maximum density of a crystal $\rho_{cp}$ to decrease with 
$\sigma$. Computer simulations\cite{Bolhuis-824,Schaertl-828} indicate that the 
limiting densities of amorphous and crystalline structures become equal at 
$\sigma \approx 0.05$. For higher polydispersities disordered structures fill 
space more efficiently than ordered ones. Consequently the appearance of an 
equilibrium amorphous phase and the ensuing re-entrant freezing transition 
should be a universal feature of all polydisperse systems -- so answering the 
last of our questions.

In conclusion we have presented a simple mean-field model of polydisperse hard 
spheres which suggests that the equilibrium state at high polydispersities and 
densities is amorphous. An equilibrium crystal is found only at intermediate 
densities. The growing stability of the fluid phase with polydispersity causes a 
singularity in the density-polydispersity phase diagram which we identify as a 
point of equal concentration. Finally, although we use mean-field theory, our 
results should be robust with respect to fluctuation effects since the 
transition we find is not  critical and the thermodynamic functions are not 
singular at this point.

It is a pleasure to thank Mike Cates, Peter Sollich and Richard Sear for helpful 
discussions. The work was made possible by financial assistance from the 
Engineering and Physical Sciences Research Council.

%
%


\begin{references}
\bibitem{foot}

In the literature $\sigma_{t}$ is commonly called the ``critical'' 
polydispersity. This choice is unfortunate because the singularity at $\sigma = 
\sigma_{t}$ is unconnected with critical phenomena.
\bibitem{note1}
The disordered phase lies on a high density continuation of the fluid branch. At 
these densities, the most likely structure is that of an amorphous glass [see, 
e.g. Ref.~\cite{Moriguchi-1232}].
\end{references}
\end{document}